# Low density flux pinning in YBCO single crystals

V.Yu.Monarkha, V.A.Pashchenko, and V.P.Timofeev

*B.Verkin Institute for Low Temperature Physics & Engineering National Academy of Sciences of Ukraine,*
*47, Lenin Ave., 61103 Kharkov, Ukraine*
*E-mail: timofeev@ilt.kharkov.ua*

**Abstract.** The Abricosov vortex and bundles dynamics was experimentally investigated in Earth's magnetic field range. Isothermal relaxation features in YBCO single crystal samples with strong pinning centers were studied for different sample-field orientation. The normalized relaxation rate $S$ obtained allowed to estimate the effective pinning potential $U$ in the bulk of the YBCO sample and its temperature dependence, as well as the critical current density $J_c$. A comparison between the data obtained and the results for similar measurements in significantly higher magnetic fields was performed. To compare different $J_c$ measuring techniques magnetization loop $M(H)$ measurements, were made. These measurements provide many important parameters of the sample under study (penetration field $H_p$, first critical field $H_{c1}$, etc.) that contain the geometrical configuration of the samples.

Experimental study of magnetic flux dynamics in high temperature superconductors (HTSC) in wide temperature and field range is important for understanding the main mechanisms of pinning and vortex creep, and while constructing superconducting electronics. Presently the low field (close to earth's magnetic field) range and the close to critical temperature range are the most unstudied [1]. The vortex hopping probability grows exponentially with the temperature increasing and the pinning force decreasing, therefore the HTSC microstructure and the corresponding activation energies play a major role in the magnetic flux dynamics [2].

In this paper we present an experimental study of low density captured magnetic flux dynamics in YBCO single crystal samples near the superconducting phase transition point ($0,5 \leq T/T_c < 0,99$). The relaxation of isothermal magnetic momentum created by captured magnetic flux (single vortexes or vortex bundles) was registered and the averaged effective pinning potential $U$ was obtained using Anderson-Kim linear thermo activated flux creep model. By comparing the results with different crystalline structure we concluded that thermo activated transformation of the Josephson weak links in the system of unidirectional twin boundaries (TB) affects the effective pinning potential. Comparative magnetization loop measurements of the single crystal samples in relatively small fields ($H_{max} \leq 300$ Oe) were performed.

As the main research object we have selected undoped orientated YBCO single crystals. Samples under study had dimensions close to 1x1 mm$^2$, and thickness from 0,015 to 0,02 mm. To analyze the role of the planar defects in pinning process and flux dynamics we have chosen single crystal YBCO samples with twinning boundaries (TB) going all along the sample's volume and parallel to the $c$ axis and with minimal mosaicity. Contactless SQUID-magnetometery measurement method used provides the necessary susceptibility ($\sim 8 \cdot 10^{-11}$ A·m$^2$), acceptable thermal stabilization (~10 mK) in the measurement chamber, and allows to completely remove all sample preparation procedures and to preserve the sample's structure intact.

To study the magnetic flux dynamics associated with bulk pinning centers in YBCO single crystals the measurements were performed in FC (Field Cooling) mode. When the sample transits to the superconducting state the major part of the magnetic field is being pushed outside the sample's volume, while the other part in the form of magnetic vortices and their boundles is being captured by various defects all over the crystal. In this case the role of the surface barriers in magnetic flux dynamics is minimal [1]. In this simplest case the effective pinning potential $U$ can be obtained from the normalized magnetic relaxation rate using Anderson-Kim's model: $S = 1/M_0 \, (dM/d\ln t) = -kT/U$, Where $M_0$ is the initial magnetization value of the sample, $k$ is the Boltzman's constant and $t$ is time. The major part of published studies on investigating the pinning mechanisms and magnetization relaxation in HTSC with different crystalline structure were performed in strong magnetic fields (~kOe), when the main role is played by the well formed vortex grid. In this case weak Josephson links are suppressed, the measurement data is extremely sensitive to field-sample orientation and to linear and planar defects. The low field area and the initial part of the magnetization curves remain less studied due to high sensitivity required and the electromagnetic interference.

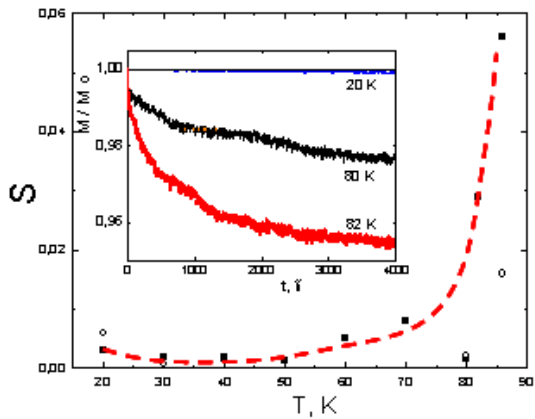

Fig.1. Averaged normalized isothermal magnetization relaxation rate (insert-typical $M$ relaxation).

As it was shown in [4], from the point of view of the collective pinning theory in weak magnetic fields the noninteracting vortex creep takes place. In this case the flux movement speed, the correlation length $L_0$ and the pinning potential are independent from the field strength $H$, and the measurement results are weakly affected by the deviation of the field induction vector from the main crystallographical axis of the YBCO crystal. Apart from that the critical current value $J_c$, defined by the equilibrium of the pinning force and Louretz force $J_c = U_c/\Phi_0 L_0$ ($\Phi_0$ – flux quantum), is not sensitive to the angle between the main cystallographical axis, twinning plane and the field induction $H$.

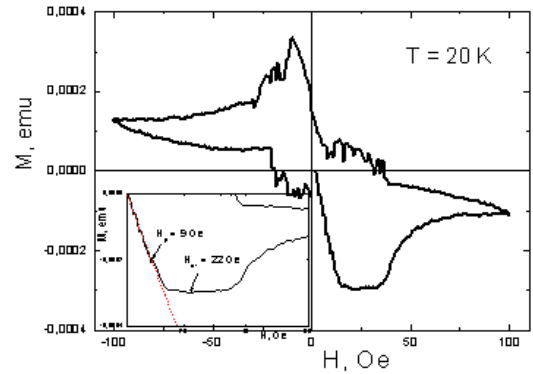

Fig.3 Magnetization loop of the sample

To obtain the superconducting phase transition curve $M(T)$ the sample was cooled in low magnetic field. Then the field was turned off and the sample was heated up at a constant speed of $dT/dt \approx 0,2$ K/min while the magnetization of the sample and it's temperature were registered. After that the magnetization relaxation measurements could be performed in the selected areas of field strengths and temperatures. The sample was cooled in the specified field. When the selected temperature of the sample became stable the field was turned off and the magnetization of the sample was registered in time.

Fig.1 shows the behavior of the averaged normalized isothermal magnetization relaxation rate of a typical sample under study at different temperatures.

The comparison of our results on the relaxation rate behavior with the results obtained in significantly higher magnetic fields by other authors [5] is presented on the Fig.2. The sample-field orientation was $H \parallel c$ at which the most effective vortex pinning was observed. The magnetization curves of the type-II superconductors containing various pinning centers become irreversible and hysteretic. For the typical YBCO samples the presence of pinning centers leads to the transformation of the $M(H)$ curve to a magnetization loop.

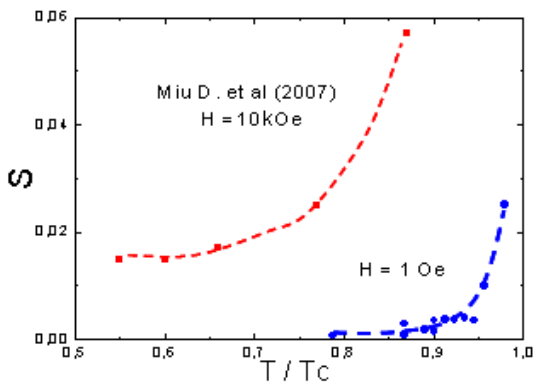

Fig.2. Comparison of the relaxation rate temperature dependance for different field strength.

The width of the magnetization loop $\Delta M$ is proportional to the averaged effective pinning potential. According to Bean's critical state model the critical current density $J_c$ is associated with the geometrical parameters and the $\Delta M$ of the sample under study. The value of $J_c$ can be estimated for example by a simple relation: $J_c = 15 \, \Delta M/R$, where $R$ − is the function of the sample's geometry which include the demagnetizing factor and the $J_c(H)$ dependence [6]. A similar relation $J_c = 20 \, \Delta M/[a(1-$

a/3b)], where a,b (a<b) – crossection dimensions of the sample, can be found in [7].

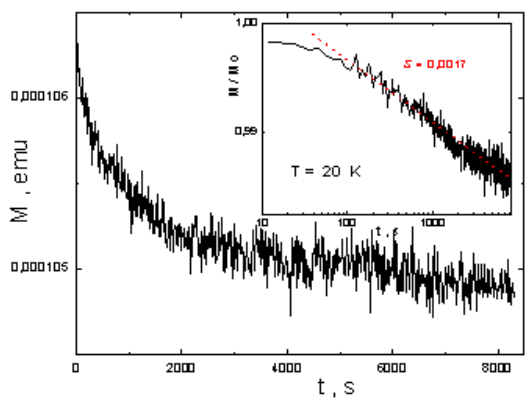

Fig.4. Magnetization relaxation curve (insert - *S* estimation method).

In order to compare the two different flux dynamics investigation methods we performed a series of measurements on MPMS 5 (Quantum Design). Fig.3 shows the typical YBCO *M(H)* curve at 20K. The initial part of the magnetization curve, shown on the insert, allows to estimate the penetration field $H_p$ and the first critical field. The magnetization loop is asymmetrical with regard to the *H* axis which indicates the strong influence of the Bin-Livingston surface barrier and the thermo activated collective creep of vortices and its bundles [8]. From the results obtained it seams not to be possible to apply the Bin's critical state model to estimate the critical current density $J_c$ in the selected field and temperature ranges. Thus the magnetization relaxation method is more applicable in this case. Fig.4 shows the magnetization relaxation conditional on the flux creep captured in 10 Oe field. These conditions correspond to the initial part of the hysteretic loops and referring to the literature is less studied.

**Conclusion**

It was shown that that the crystalline structure (mostly unidirectional twin boundaries) of the YBCO single crystals has significant influence on the magnetization relaxation rate, at constant low fields ($H \approx$ 1 - 10 Oe) in wide temperature range close to $T_c$.

The results obtained allow to reevaluate the magnetic flux dynamics defined by the significant growth of the effective pinning potential in low flux density case. From a practical point of view the results of our study can be useful while constructing high sensitive superconducting devices, to minimize self-noises of the HTSC sensors and to increase sensitivity of the liquid nitrogen cooled apparatus.